\documentclass[letterpaper,twoside,12pt]{article}
\usepackage{amssymb}
\usepackage{latexsym}
\usepackage{verbatim}
\usepackage{floatfig,epsfig}
\usepackage{calc}
\usepackage{amstext,amsfonts}
\usepackage{fancyheadings}
\usepackage{pstricks,pst-node}
\usepackage{rotating,graphicx}
\newlength{\treescale}
\newlength{\treeunit}
\newcommand{\bfC}{\mathbf{C}}
\newcommand{\papertitle}{%
Normal Coordinates  and Primitive Elements\\[3mm] 
in the Hopf Algebra of Renormalization%
}
\newcommand{\headtitle}{%
Normal Coordinates  and Primitive Elements...%
}
\newcommand{\paperauthor}{%
C.{} Chryssomalakos, H.{} Quevedo, 
M.{} Rosenbaum and J.{} D.{} Vergara%
}
\newlength{\enviropost}
\newcommand{\be}{\begin{equation}}
\newcommand{\ee}{\end{equation}}
\newcommand{\ble}[1]{\begin{equation} \label{#1}}
\newcommand{\bae}{\begin{eqnarray}}
\newcommand{\eae}{\end{eqnarray}}
\newcommand{\fle}[2]%
{\vspace{1.5ex}
\be
\label{#1}
\mbox{%
\setlength{\fboxsep}{3ex}%
\framebox{$\dss #2 $}}
\ee} 

\newcommand{\nn}{\nonumber}
\newcommand{\ff}{\nn \\}
\newcommand{\fe}{& = &}
\newcommand{\aqqa}{& \qquad &}

\newtheorem{theorem}{Theorem}

\newtheorem{lemma}{Lemma}
\newtheorem{exatitle}{Example}
\newenvironment{proof}%
{\noindent \textsc{Proof}:\ }%
{\hfill $\blacksquare$  \vspace{\enviropost} \\}
\newenvironment{example}[2]%
{\begin{exatitle} \label{#2} #1 \end{exatitle}}%
{\hfill $\Box$ \vspace{\enviropost} \\}
{\vspace{3mm} \noindent {\bf Aside:} \small}%
{\hfill \rule[-3mm]{0mm}{0mm}$\Diamond$\\}
\newcommand{\scr}{\scriptscriptstyle}
\newcommand{\dss}{\displaystyle}
\newcommand{\id}{\mathop{\rm id}}
\newcommand{\ra}{\rightarrow}

\newcommand{\ot}{\otimes}
\newcommand{\tr}{\triangleright}
\newcommand{\ip}[2]{\left\langle #1, #2\right\rangle}

\newcommand{\uA}{1_\calA}

\newcommand{\rsub}{\rule{0ex}{2ex}} 
\newcommand{\ibf}{\mathbf{i}}
\newcommand{\eg}{\hbox{\it e.g.{}}}

\newcommand{\ie}{\hbox{\it i.e.{}}}

\newcommand{\rhs}{\hbox{r.h.s.{}}}
\newcommand{\cf}{\hbox{cf.{}}}
\newcommand{\longpage}{\enlargethispage{\baselineskip}}

\newcommand{\calA}{\mathcal{A}}

\newcommand{\calF}{\mathcal{F}}
\newcommand{\calG}{\mathcal{G}}
\newcommand{\calH}{\mathcal{H}}

\newcommand{\calO}{\mathcal{O}}

\newcommand{\calR}{\mathcal{R}}

\newcommand{\calT}{\mathcal{T}}
\newcommand{\calU}{\mathcal{U}}

\newcommand{\slte}{\setlength{\treescale}{1.8\treeunit}}
\newcommand{\sltt}{\setlength{\treescale}{1.5\treeunit}}
\newcommand{\slts}{\setlength{\treescale}{1.2\treeunit}}
\newcommand{\sltss}{\setlength{\treescale}{1.2\treeunit}}
\newcommand{\n}[2]{{n_{\rsub #1}}^{\! \!  #2}}
\newcommand{\f}[2]{{f_{\rsub #1}}^{\! \!  #2}}
\newcommand{\Dlin}{\Delta_{\mbox{\em \scriptsize lin}}}
\newcommand{\psilin}{\psi_{\mbox{\em \scriptsize lin}}}
\begin{document}
\normalsize
\initfloatingfigs
\begin{titlepage}
\vspace*{-1cm}
\begin{flushright}
\textsf{ICN-UNAM-01/08}
\\
\mbox{}
\\
\textsf{May 17, 2001}
\\[3cm]
\end{flushright}
\renewcommand{\thefootnote}{\fnsymbol{footnote}}
\begin{LARGE}
\bfseries{\sffamily \papertitle}
\end{LARGE}

\noindent \rule{\textwidth}{.6mm}

\vspace*{1.6cm}

\noindent 
\begin{large}%
\textsf{\bfseries%
\paperauthor
}
\end{large}


\phantom{XX}
\begin{minipage}{.8\textwidth}
\begin{it}
\noindent Instituto de Ciencias Nucleares \\
Universidad Nacional Aut\'onoma de M\'exico\\
Apdo. Postal 70-543, 04510 M\'exico, D.F., MEXICO \\
\end{it}
\texttt{chryss,quevedo,mrosen,vergara@nuclecu.unam.mx
\phantom{X}}
\end{minipage}
\\

\vspace*{3cm}
\noindent
\textsc{\large Abstract: }
We introduce normal coordinates on the infinite dimensional group
$G$ introduced by Connes and Kreimer in their analysis of the Hopf
algebra of rooted trees. 
We study the primitive elements of the algebra and show that they
are generated by a
simple application of the inverse Poincar\'e lemma,
given a closed left invariant 1-form on $G$.
For the special case of the ladder primitives, we find a second
description that relates them to
the Hopf algebra of functionals on power series with the usual
product.
Either approach shows that the ladder primitives are given by the 
Schur polynomials. 
The relevance of the lower central series of the dual Lie algebra
in the process of renormalization is also discussed, leading to a
natural concept of $k$-primitiveness, which is shown to be 
equivalent to the one already in the literature.
\end{titlepage}
\setcounter{footnote}{1}
\renewcommand{\thefootnote}{\arabic{footnote}}
\setcounter{page}{2}
\noindent \rule{\textwidth}{.5mm}

\tableofcontents

\noindent \rule{\textwidth}{.5mm}
\newsavebox{\Treeoo}
\newsavebox{\Treeto}
\newsavebox{\Treetho}
\newsavebox{\Treetht}
\newsavebox{\Treefo}
\newsavebox{\Treeft}
\newsavebox{\Treefth}
\newsavebox{\Treeff}
\newsavebox{\Treefio}
\newsavebox{\Treefit}
\newsavebox{\Treefith}
\newsavebox{\Treefif}
\newsavebox{\Treefifi}
\newsavebox{\Treefis}
\newsavebox{\Treefise}
\newsavebox{\Treefie}
\newsavebox{\Treefin}
\sbox{\Treeoo}{{%
\begin{pspicture}(2mm,-1mm)(8mm,10mm)
\psset{xunit=1cm,yunit=1cm}
\psdots[dotscale=3](.5,0)
\end{pspicture}}%
}
\newcommand{\Too}[1][0]{%
\raisebox{#1\totalheight}%
  {\resizebox{!}{\treescale}{\usebox{\Treeoo}}}}
\sbox{\Treeto}{{%
\begin{pspicture}(2mm,-1mm)(8mm,10mm)
\psset{xunit=1cm,yunit=1cm}
\psline[linewidth=.3mm]{*-*}%
(.5,1)(.5,0)
\psdots[dotscale=3](.5,1)(.5,0)
\end{pspicture}}%
}
\newcommand{\Tto}[1][0]{%
\raisebox{#1\totalheight}%
  {\resizebox{!}{\treescale}{\usebox{\Treeto}}}}
\sbox{\Treetho}{{%
\begin{pspicture}(2mm,-1mm)(8mm,2cm)
\psset{xunit=1cm,yunit=1cm}
\psline[linewidth=.3mm]{*-*}%
(.5,2)(.5,0)
\psdots[dotscale=3](.5,2)(.5,1)(.5,0)
\end{pspicture}}%
}
\newcommand{\Ttho}[1][0]{%
\raisebox{#1\totalheight}%
  {\resizebox{!}{2\treescale}{\usebox{\Treetho}}}}
\sbox{\Treetht}{{%
\begin{pspicture}(-3mm,-1mm)(13mm,10mm)
\psset{xunit=1cm,yunit=1cm}
\psline[linewidth=.3mm]{*-*}%
(.5,1)(0,0)
\psline[linewidth=.3mm]{*-*}%
(.5,1)(1,0)
\psdots[dotscale=3](.5,1)(0,0)(1,0)
\end{pspicture}}%
}
\newcommand{\Ttht}[1][0]{%
\raisebox{#1\totalheight}%
  {\resizebox{!}{\treescale}{\usebox{\Treetht}}}}
\sbox{\Treefo}{{%
\begin{pspicture}(2mm,-1mm)(8mm,30mm)
\psset{xunit=1cm,yunit=1cm}
\psline[linewidth=.3mm]{*-*}%
(.5,3)(.5,0)
\psdots[dotscale=3](.5,3)(.5,2)(.5,1)(.5,0)
\end{pspicture}}%
}
\newcommand{\Tfo}[1][0]{%
\raisebox{#1\totalheight}%
  {\resizebox{!}{3\treescale}{\usebox{\Treefo}}}}
\sbox{\Treefth}{{%
\begin{pspicture}(2mm,-1mm)(13mm,20mm)
\psset{xunit=1cm,yunit=1cm}
\psline[linewidth=.3mm]{*-*}%
(.5,2)(.5,0)
\psline[linewidth=.3mm]{*-*}%
(.5,2)(1,1)
\psdots[dotscale=3](.5,2)(.5,1)(.5,0)(1,1)
\end{pspicture}}%
}
\newcommand{\Tfth}[1][0]{%
\raisebox{#1\totalheight}%
  {\resizebox{!}{2\treescale}{\usebox{\Treefth}}}}
\sbox{\Treeft}{{%
\begin{pspicture}(-3mm,-1mm)(13mm,20mm)
\psset{xunit=1cm,yunit=1cm}
\psline[linewidth=.3mm]{*-*}%
(.5,2)(.5,1)
\psline[linewidth=.3mm]{*-*}%
(.5,1)(0,0)
\psline[linewidth=.3mm]{*-*}%
(.5,1)(1,0)
\psdots[dotscale=3](.5,2)(.5,1)(0,0)(1,0)
\end{pspicture}}%
}
\newcommand{\Tft}[1][0]{%
\raisebox{#1\totalheight}%
  {\resizebox{!}{2\treescale}{\usebox{\Treeft}}}}
\sbox{\Treeff}{{%
\begin{pspicture}(-3mm,-1mm)(13mm,10mm)
\psset{xunit=1cm,yunit=1cm}
\psline[linewidth=.3mm]{*-*}%
(.5,1)(0,0)
\psline[linewidth=.3mm]{*-*}%
(.5,1)(.5,0)
\psline[linewidth=.3mm]{*-*}%
(.5,1)(1,0)
\psdots[dotscale=3](.5,1)(0,0)(.5,0)(1,0)
\end{pspicture}}%
}
\newcommand{\Tff}[1][0]{%
\raisebox{#1\totalheight}%
  {\resizebox{!}{\treescale}{\usebox{\Treeff}}}}
\sbox{\Treefio}{{%
\begin{pspicture}(2mm,-1mm)(8mm,40mm)
\psset{xunit=1cm,yunit=1cm}
\psline[linewidth=.3mm]{*-*}%
(.5,4)(.5,0)
\psdots[dotscale=3](.5,4)(.5,3)(.5,2)(.5,1)(.5,0)
\end{pspicture}}%
}
\newcommand{\Tfio}[1][0]{%
\raisebox{#1\totalheight}%
  {\resizebox{!}{\treescale}{\usebox{\Treefio}}}}
\section{Introduction}
\label{Intro}
The process of renormalization in quantum field theory has been
substantially elucidated in recent years. In a series of papers
(see, \eg,~\cite{Kre:99,Con.Kre:00,Bro.Kre:99,Kre:00} and references
therein),
a Hopf algebra structure has been identified that
greatly simplifies its combinatorics. This, in turn, has led to
the development of an underlying geometric picture, involving
an infinite dimensional group manifold
$G$, the
coordinates of which are in one-to-one correspondence with
(classes of) 
1PI superficially divergent Feynman diagrams of the theory. The latter
are indexed by a type of graphs known as (decorated) rooted trees,
which capture the subdivergence structure of the
diagram. The forest formula prescription for the renormalization
of a diagram then is 
translated into
a series of operations on the corresponding rooted tree and the
latter  have been shown to deliver 
standard Hopf algebraic quantities, like the coproduct and the
antipode of the rooted tree. The above results were obtained
using
a powerful mixture of algebraic and combinatoric techniques that
brought to light unexpected interconnections with noncommutative
geometry, among several
other fields.  

The complexity of the full Hopf algebra of decorated rooted trees
is, in many respects,
overwhelming. Even in the simplest cases, one is
confronted with an infinite set of available decorations for the
vertices of the rooted trees, originating in the infinite number of
primitive divergent diagrams appearing in the underlying theory. 
It is rather
fortunate then that the considerably simpler algebra of rooted
trees with a single decoration seems to capture many of the
features of realistic theories. It is for this reason that it has
been studied extensively, as a first step towards an
understanding of the full theory. Of primary importance, given
their r\^ole in renormalization theory, is the study of the
primitive elements of the above Hopf algebra. These correspond to sums
of products of diagrams with the property that their
renormalization involves a single subtraction. In
Ref.~\cite{Bro.Kre:00}, an ansatz is presented for a
(conjectured) infinite  family of such elements, corresponding to
the ladder generators of the algebra, \ie, to trees whose every
vertex has fertility at most one. Furthermore, dealing with the
general case, a set of vertex-increasing operators
is constructed, that generates new primitive elements from known
ones. As the number of primitive elements increases rapidly with
increasing number of vertices, this approach necessitates the
introduction of new operators in each step, a task that has not
yet been systematized. 

Our motivation in this paper is two-fold. On a general, 
methodological level, we
argue that the above algebraic/combinatoric approach, with all
its multiple successes, 
should nevertheless be complemented by a
differential geometric one, which, we feel, has not been
sufficiently considered in
the literature. On a second, more concrete level, we provide
support for our claim, by showing how a simple application of the
inverse Poincar\'e lemma reduces the search for primitive
elements to that of closed, left invariant (LI) 1-forms on $G$.
For the case of the ladder primitives, we give a simple
generating formula that identifies them with the Schur
polynomials. Our discussion uses the normal coordinates on
the group, a choice that leads naturally to a concept of
$k$-primitiveness, associated with the lower central series of the 
dual Lie algebra --- we prove that this coincides with 
the $k$-primitiveness introduced in Ref.~\cite{Bro.Kre:00}. 
We discuss the r\^ole of the new coordinates in
renormalization, using the toy model realization of
Ref.~\cite{Kre:98}, while also commenting on similar results obtained
for the more realistic heavy quark model of~\cite{Bro.Kre:99}. 
\section{Differential geometry \'a la Hopf}
\label{DgH}
We will be dealing with differential geometric concepts
expressed in Hopf algebraic terms. We opt for this formulation
having in mind the transcription of our results for the
non-commutative case --- Hopf algebras are ideally suited to this
task. We start by providing a short dictionary between the two
languages and establish the notation, assuming nevertheless 
familiarity with the basic definitions. 

Two algebras will be of main interest to us: on the one hand we
have the (commutative, non-cocommutative) algebra $\calA$ of 
functions on a (possibly infinite dimensional)
group manifold, generated by $\{ \phi^A \}$, with $A$ ranging in
an index set - we denote by $a$, $b, \ldots$ general elements of
$\calA$. 
On the other hand, we have the (non-commutative,
cocommutative) universal enveloping algebra 
$\calU$ of the Lie algebra of the group. We actually work with a
suitable completion of $\calU$, so as to allow exponentials of its
generators $Z_A$, which we identify with the points of the
manifold%
\footnote{%
The particular group we deal with in Sect.~\ref{Hartd} is non-compact
and infinite dimensional. Nevertheless, in this paper, we only  
consider elements
that correspond to exponentials of linear combinations of
the generators. For a readable account of what we 
might be missing in doing so, see Ref.~\cite{Mil:84}%
}
- we denote by $x$, $y , \ldots$ general elements of
$\calU$ (we use  $g$, $g' , \ldots$ if we refer to group
elements in particular).

Both algebras are Hopf algebras. For $\calA$, the
{\em coproduct} 
$\Delta(a) \equiv
a_{(1)} \ot a_{(2)}$ codifies left
and right translations 
\ble{Ltrc}
L_g^*(a)(\cdot) = a_{(1)}(g) a_{(2)}(\cdot)
\, ,
\ee
and similarly for right translations. For
$\calU$, it expresses Leibniz's rule, $\Delta(Z) = Z \ot 1 + 1 \ot
Z$, for the left-invariant generator $Z$. The two Hopf algebras 
are {\em dual} , via the {\em inner product} (also called {\em
pairing})
\ble{ipdef}
\ip{\cdot}{\cdot}: \calU \ot \calA \ra \mathbb{C}
\, ,
\qquad
x \ot a \mapsto \ip{x}{a}
\, ,
\ee
which, when $x$ stands for a generator $Z$, amounts to
taking the derivative of $a$ along $x$
and evaluating it at the identity. For $x=g$, the above
definition produces a Taylor series expansion of $a$ at the
identity which gives, for $a$ analytic, the value $a(g)$ of $a$ 
at the point $g$. The coproduct in $\calA$ is dual to the product in
$\calU$ via
\ble{pdc}
\ip{xy}{a} = \ip{x \ot y}{a_{(1)} \ot a_{(2)}}
\ee
and {\em vice-versa}.
We usually work with
{\em dual bases}, so that $Z_A$ only gives 1 when paired with $\phi^A$,
while its inner product with all other $\phi$'s, as well as with
all products of $\phi$'s, vanishes. Given a
Poincar\'e-Birkhoff-Witt basis $\{f^i\}$ for $\calA$, 
\ble{PBdWA}
\{ f^i\} = \{1, \phi^A, \phi^A \phi^B, \ldots\}
\, ,
\ee
one can build a dual basis
$\{e_i\}$ for the entire $\calU$ by adjoining to the above $Z$'s
polynomials in them, $\{e_i\} = \{1, Z_A,$ quadratic, cubic,
$\ldots \}$, with $\ip{e_i}{f^j} = \delta_i^j$ --- this, in
general, involves a non-trivial calculation. 

To every element $a$ of $\calA$ we can associate a LI 1-form 
$\Pi_a$, given by 
\ble{Pia}
\Pi_a = S(a_{(1)}) d a_{(2)}
\, ,
\ee
$d$ being the
exterior derivative and $S$ the antipode in $\calA$. $\Pi$ is
linear, while on products it gives 
\ble{Pipro}
\Pi_{(ab)} = \Pi_a \epsilon(b) + \epsilon(a) \Pi_b  
\, ,
\qquad \qquad
\mbox{$\calA$ commutative}
\, ,
\ee
where $\epsilon$ is the counit in $\calA$. We take all generators
$\phi^T$ of $\calA$ to be couniteless, \ie, we choose functions that
vanish at the identity of the group, except for the unit function
$\uA$ (which we often write as just 1). This implies that $\Pi$
only returns a non-zero result when applied to the generators and
vanishes on all products, as well as on $\uA$. The Maurer-Cartan 
(MC) equations take the form
\ble{MCeq}
d \Pi_a = - \Pi_{a_{(1)}} \Pi_{a_{(2)}} 
\, .
\ee
Using~(\ref{Pipro}), one sees that only the bilinear part of the
coproduct contributes to the MC equations. 
\section{The Hopf Algebra of Rooted Trees and its Dual}
\label{Hartd}
\subsection{Functions}
\label{functions}
We specialize the general considerations of the previous section
to the Connes-Kreimer algebra of renormalization. For a detailed 
exposition we refer the
reader to~\cite{Kre:98,Con.Kre:98,Kas:01} and references therein, 
we give here only a brief account of the
basic definitions and some illustrative examples. 
$\calA$ is now the Hopf algebra $\calH_R$ of functions
generated by $\phi^T$, where $T$ is a rooted tree. This means
that the group manifold $G$ is, in this case, infinite dimensional,
with one dimension for every rooted tree - the $\phi$'s are 
coordinate functions on this manifold. The group law is encoded in
the coproduct
\ble{copA}
\Delta(\phi^T) = \phi^T \ot 1 + 1 \ot \phi^T + 
\sum_{\mbox{\small cuts } C} 
\phi^{P^C(T)}  \ot \phi^{R^C(T)}
\, .
\ee
The sum in the above definition is over {\em admissible} cuts,
\ie, cuts that may involve more than one edge ({\em simple} cuts)
but such that there is no more than one simple cut on any
path from the root downwards. $R^C(T)$ is the part that is left
containing the root of $T$ while $P^C(T)$ is the product of all
branches cut, \eg
\slte
\ble{excopA}
\Delta(\Ttht) = \Ttht \ot 1 + 1 \ot \Ttht + 2 \Too \ot \Tto +
\Too \Too \ot \Too
\, ,
\ee
where we let a tree $T$ itself denote the corresponding function
$\phi^T$, a
convention freely used in the rest of the paper.
The factor 2 on the r.h.s. appears because there are two
possible cuts on $\Ttht$ generating the corresponding term. 
A convenient way to recast~(\ref{copA}) as a single sum, is to
introduce a {\em full} and an {\em empty cut}, above and below
any tree $T$ respectively, \eg,
\ble{fec}
\begin{pspicture}(-3mm,-1mm)(13mm,10mm)
\setlength{\unitlength}{5mm}
\psset{xunit=5mm,yunit=5mm}
\psline[linewidth=.3mm]{*-*}%
(.5,1)(0,0)
\psline[linewidth=.3mm]{*-*}%
(.5,1)(1,0)
\psdots[dotscale=1](.5,1)(0,0)(1,0)
\psline[linewidth=.2mm, linestyle=dotted]{-}%
(0,1.5)(1,1.5)
\put(1.4,1.35){\makebox[0mm][l]{full cut}}
\end{pspicture}
\qquad 
\qquad 
\qquad 
\begin{pspicture}(-3mm,-1mm)(13mm,10mm)
\setlength{\unitlength}{5mm}
\psset{xunit=5mm,yunit=5mm}
\psline[linewidth=.3mm]{*-*}%
(.5,1)(0,0)
\psline[linewidth=.3mm]{*-*}%
(.5,1)(1,0)
\psdots[dotscale=1](.5,1)(0,0)(1,0)
\psline[linewidth=.2mm, linestyle=dotted]{-}%
(0,-.5)(1,-.5)
\put(1.4,-.7){\makebox[0mm][l]{empty cut}}
\end{pspicture}
\ee
We rewrite~(\ref{copA}) in the form
\ble{copAp}
\Delta(\phi^T) =  
\sum_{\mbox{\small cuts } C'} 
\phi^{P^{C'}(T)}  \ot \phi^{R^{C'}(T)}
\, ,%
\ee
where the above two extra cuts, included in $C'$, produce the primitive
part of the coproduct.
Notice that $\Delta$ respects the grading given by the
number $v(T)$  of vertices of a tree $T$. We call this the 
$v$-degree of $\phi^T$,
denote it by $\mbox{deg}_v(\phi^T)$, and extend it to monomials 
as the sum of the $v$-degrees of
the factors. The polynomial degree will be called $p$-degree to
avoid confusion --- it is obviously not respected by the
coproduct. We will use the notation $\calA^{(n)}_i$ for the
subspace of $\calA$ of $v$-degree $n$ and $p$-degree $i$, \eg,
$\calA^{(n)}_1$ is the linear span of the generators with $n$
vertices. 

\subsection{Vector fields}
\label{vectorf}
The r\^ole of $\calU$ is now played by $\calH_R^*$, generated by 
$\{ Z_T \}$, with $T$ a rooted tree and we
take the $Z$'s dual to the $\phi$'s, in the sense of the previous
section. $Z_T$ is a left invariant vector field on $G$.  The
Lie algebra of such vector fields is found by computing,
using~(\ref{pdc}),
the pairing of $Z_A Z_B - Z_B Z_A$ with all basis functions 
$\{ f^i \}$. 
\begin{example}{Computation of $[ \slts Z_{\Too}, \, Z_{\Tto}
]$}{ZZcomex}
We have
\slte
\ble{copfun}
\tilde{\Delta}(\Ttho) = \Too \ot \Tto + \Tto \ot \Too
\, ,
\qquad
\tilde{\Delta}(\Ttht) = 2 \, \Too \ot \Tto + \Too \Too \ot \Too
\, ,
\qquad
\tilde{\Delta}(\Too \Tto) = \Too \ot \Tto + \Tto \ot \Too
                + \Too \Too \ot \Too + \Too \ot \Too \Too
\, ,
\ee
where $\tilde{\Delta}(\phi^T) \equiv \Delta(\phi^T) - \phi^T \ot
1 - 1 \ot \phi^T$. 
These are the only functions that contain the term $\sltt \Too \ot
\Tto$ in their coproduct. We find therefore, using~(\ref{pdc}),
\ble{ipZZf1}
\ip{\slts Z_{\Too} Z_{\Tto}}{\slte  \Ttho} = 1
\, ,
\qquad
\ip{\slts Z_{\Too} Z_{\Tto}}{\slte  \Ttht} = 2
\, ,
\qquad
\ip{\slts Z_{\Too} Z_{\Tto}}{\slte \Too \Tto} = 1
\, .
\ee
Similarly, one computes
\ble{ipZZf2}
\ip{\slts Z_{\Tto} Z_{\Too}}{\slte \Ttho} = 1
\, ,
\qquad
\qquad
\ip{\slts Z_{\Tto} Z_{\Too}}{\slte \Too \Tto} = 1
\, ,
\ee
the pairings with all other functions being zero. It follows that
the only non-zero pairing of the commutator is
\ble{ipcom}
\ip{[Z_{\slts \Too}, \, Z_{\slts \Tto} \, ]}{\Ttht} = 2
\, .
\ee
But the element $2 Z_{\slts \Ttht}$ of $\calU$ has  
exactly the same pairings, therefore, in order
for  the
inner product between $\calU$ and $\calA$ to be non-degenerate, one must
set $[ Z_{\slts \Too}, \, Z_{\slts \Tto} ] = 2 Z_{\slts \Ttht}$.
\end{example}
\noindent Proceeding along these lines, one arrives at the
general expression~\cite{Con.Kre:00}
\ble{ZZpr}
[ Z_{T_1}, \,   Z_{T_2}] 
= \sum_{T} 
\left( \n{T_1 T_2}{T} - \n{T_2 T_1}{T} \right) \, Z_T
\equiv 
\sum_{T}
\f{T_1 T_2}{T}  \, Z_T
\, ,
\ee
where $\n{T_1 T_2}{T}$ is the number of simple cuts on $T$ that
produce $T_1$, $T_2$, with $T_2$ containing the root of $T$
(denoted by $n(T_1,T_2,T)$ in~\cite{Con.Kre:98}) and the last
equation defines the structure constants $\f{T_1 T_2}{T}$ of the
Lie algebra. We introduce, following~\cite{Con.Kre:00}, a
$*$-operation among the $Z$'s, defined by
\ble{stardef}
Z_{T_1} * Z_{T_2} = \n{T_1 T_2}{T} Z_T
\, .
\ee
Notice that this is {\em not} the product in $\calU$ but,
nevertheless, it gives correctly the commutator when
antisymmetrized (\cf~(\ref{ZZpr})).
The above Lie bracket conserves the number of vertices. 
\subsection{1-forms}
\label{forms}
We turn now to LI 1-forms. Starting
from~(\ref{Pia}) and using the particular form of the coproduct
in~(\ref{copA}), we find
\fle{LIform}{%
\Pi_{\phi^T} 
\, = \, 
\sum_{C'} \phi^{S(P^{C'}(T))} \, d \phi^{R^{C'}(T)}
\, = \, 
d \phi^T 
+ \sum_{C} \phi^{S(P^C(T))} \, d \phi^{R^C(T)}
\, .%
}
For the MC equations we may use directly~(\ref{MCeq}) and the
comment that follows it to find
\fle{MCtree}{%
d \Pi_{\phi^T} = - \sum_{\mbox{\small simple } C} \Pi_{\phi^{P^C(T)}}
\, \Pi_{\phi^{R^C(T)}}
\, .%
}
The restriction to simple cuts is possible since cuts that
involve more than one edge produce non-linear terms in the first
tensor factor of the coproduct and these are annihilated by
$\Pi$. This is probably the easiest way to derive the structure
constants. 
\begin{example}{Maurer-Cartan equation for $\Pi_{\slts \Ttht}$}{MCP}
Using~(\ref{LIform}) we find
\ble{Piof}
\slte
\Pi_{\slts \Too} = d \Too
\, ,
\qquad
\Pi_{\slts \Tto} = d \Tto - \Too d \Too
\, ,
\qquad
\Pi_{\slts \Ttht} = d \Ttht -2 \, \Too d \Tto 
                        + \Too \Too d \Too
\, .
\ee
Direct application of $d$ to the above expression for 
$\Pi_{\slts \Ttht}$,
or use of~(\ref{MCtree}), gives
\slts
\ble{MCTtht}
d \Pi_{\Ttht} = -2 \, \Pi_{\Too} \Pi_{\Tto}
\, ,
\ee
in agreement with the commutator $[ Z_{\Too}, \, Z_{\Tto} ]=
2 Z_{\Ttht}$ of Ex.{}~\ref{ZZcomex}.
\end{example}
General vector and 1-form fields are obtained as linear
combinations of the above, with coefficients in $\calA$.
\section{Normal coordinates}
\label{Ncoo}
\subsection{A New Basis}
\label{NBHR}
We introduce new coordinates $\{ \psi^A \}$ on $G$, defined by
\ble{psidef}
\ip{g}{\psi^A} = \alpha^A
\, ,
\qquad
\mbox{where}
\quad
g=e^{\alpha^A Z_A}
\, ,
\ee
\ie, the $\psi$'s are normal coordinates centered at the origin
and, like the $\phi$'s, are indexed by rooted trees.
Of fundamental
importance in the sequel will be the {\em canonical element}
$\bfC$ (see, \eg,~\cite{Chr.Sch.Wat:93}),
given by
\ble{caneldef}
\bfC \, = \,  e_i \ot f^i \, = \, e^{Z_A \ot \psi^A} 
\, .
\ee
$\{ e_i \}$ and $\{ f^i \}$ above are dual bases  of $\calU$ and
$\calA$ respectively (see~(\ref{PBdWA})). 
In contrast with~(\ref{PBdWA}), we fix now
the $\{ e_i \}$ to be $\{ 1, \, Z_A, \, Z_A Z_B, \, \dots \}$ and
define the $\psi$'s by the second equality above
(the tensor product sign ensures that the
$Z$'s do not act on the $\psi$'s). $\bfC$ may be regarded as an
``indefinite group element" --- when the $\psi$'s get evaluated
on some specific point $g_0$ of the group manifold, $\bfC$ 
becomes $g_0$. One may also view $\bfC$ as an ``indefinite function"
on the group --- when the $Z$'s get evaluated on some particular
(analytic) $\phi_0$, the resulting Taylor series delivers $\phi_0$, 
\ie,
\ble{gprop}
\ip{e^{Z_A \ot \psi^A}}{\id \ot g_0} = g_0
\, , 
\qquad \qquad
\ip{e^{Z_A \ot \psi^A}}{\phi_0 \ot \id} = \phi_0
\, .
\ee
In the above, $g_0$, $\phi_0$ stand for {\em any} element in the 
corresponding
universal enveloping algebra, not just the generators. The second
of~(\ref{gprop}) gives the relation between the two linear bases
$\{ f^i_{(\phi)}\}$ and $\{ f^i_{(\psi)} \} $, generated by the
$\phi$'s and the $\psi$'s respectively. Indeed, taking
$\phi_0=\phi^A$ and expanding the exponential we find
\ble{expexp}
\phi^A 
\, = \, 
\sum_{m=0}^\infty \frac{1}{m!} 
\ip{Z_{B_1} \ldots Z_{B_m}}{\phi^A}
\psi^{B_1} \ldots \psi^{B_m}
\, = \, 
\psi^A +\frac{1}{2} 
\ip{Z_{B_1} Z_{B_2}}{\phi^A} 
\psi^{B_1} \psi^{B_2} 
+ \dots
\ee
\begin{lemma}
\label{chbas}
The change of linear basis in $\calA$ generated by~(\ref{expexp})
is invertible.
\end{lemma}
\begin{proof}
Notice that the linear part of $\phi^A(\psi)$ is $\psi^A$ and
also, that the above expansion preserves the $v$-degree. We 
choose a linear basis in $\calA$ with the following ordering 
\ble{bord}
\slts 
\{ 
\underbrace{ \, \, \phi^{\Too}}_{v=1}, 
\, \,
\underbrace{\phi^{\Tto}, \, \phi^{\Too} \phi^{\Too}}_{v=2},
\, \,
\underbrace{\phi^{\Ttho}, \, \phi^{\Ttht}, \, \phi^{\Tto} \phi^{\Too},
\, (\phi^{\Too})^3}_{v=3}, \ldots 
\}
\ee
namely, in blocks of increasing $v$-degree and, within each
block, non-decreasing $p$-degree. The above remarks then show
that the matrix $A$, defined by
\ble{Adef}
f^i_{(\phi)} = {A^i}_j f^j_{(\psi)}
\, ,
\ee
where $ \{ f^i_{(\psi)} \}$ is also ordered as in~(\ref{bord}), 
is upper triangular, with units along
the diagonal and hence invertible. 
\end{proof}
Notice that $A$ is in
block-diagonal form, with each block $A_v$ acting on 
$\calA^{(v)}$, $v=1,2,\ldots$ The computation of $\phi^A(\psi)$,
via~(\ref{expexp}), reduces essentially to the evaluation of the
inner product of $\phi^A$ with monomials in the $Z$'s --- this
is facilitated by the following
\begin{lemma}
\label{ipZp}
The inner product $\ip{Z_{B_1} \ldots Z_{B_m}}{\phi^A}$ is given by
\ble{ipZm}
\ip{Z_{B_1} \ldots Z_{B_m}}{\phi^A} 
= 
\ip{Z_{B_1}*  \ldots *Z_{B_m}}{\phi^A}
=
\n{B_1 \dots B_m}{A}
\, ,
\ee
where 
\ble{nTTdef}
\n{B_1 \dots B_m}{A} 
= 
\n{B_1 R_1}{A} \, \n{B_2 R_2}{R_1} 
\dots \n{B_{m-1} B_m}{R_{m-2}}
\ee
($Z_{B_1}*  \ldots *Z_{B_m}$ above is computed starting from the
right, \eg, $Z_{B_1}*Z_{B_2}*Z_{B_3} \equiv
Z_{B_1}*(Z_{B_2}*Z_{B_3}))$.
\end{lemma}
\begin{proof}
We have
\ble{ipZphi}
\ip{Z_{B_1} \ldots Z_{B_m}}{\phi^A}
=
\ip{Z_{B_1} \ot  \ldots \ot Z_{B_m}}{\Delta^{m-1}(\phi^A)}
\, .
\ee
In the above inner product, only the $m$-linear terms in
$\Delta^{m-1}(\phi^A)$ contribute, since the $Z$'s vanish on
products and the unit function.
One particular way of evaluating the $(m-1)$-fold coproduct is to
apply $\Delta$ always on the rightmost tensor factor. It is then
clear that, in this case, we may instead apply $\Dlin$, since 
$ \Bigl(
\prod_{j=1}^m \bigl( \id^{\ot \, j-1} \ot \Delta \bigr) 
\Bigr) (\phi^A)$
and 
$ \Bigl( 
\prod_{j=1}^m \bigl( \id^{\ot \, j-1} \ot \Dlin \bigr) 
\Bigr) (\phi^A)$
only differ by terms containing products of the $\phi$'s or
units (this is only true if $\Dlin$ is applied in the rightmost
factor).
Notice now that the $*$-product of the $Z$'s is dual to $\Dlin$
\ble{stardu}
\ip{Z_{B_1} * Z_{B_2}}{\phi^A} 
= 
\ip{Z_{B_1} \ot Z_{B_2}}{\Dlin(\phi^A)}
\, .
\ee
Repeated application of this equation and use of the definition
of $*$, Eq.~(\ref{stardef}), completes the proof.
\end{proof}
A concise way to express the relation between the two sets of
generators is via the {\em $*$-exponential} ($x \in \calU_1$)
\ble{starexp}
e_*^x \equiv \sum_{i=0}^\infty \frac{1}{i!} x^{*i} 
= \sum_{i=0}^\infty \frac{1}{i!} 
\underbrace{x * \dots *x}_{\mbox{$i$ factors}}
\, .
\ee
Combining~(\ref{expexp}) and~(\ref{ipZm}) we find
\ble{ppstar}
e_*^{Z_A \ot \psi^A} = Z_B \ot \phi^B
\, ,
\ee
where the convention $(Z_A \ot \psi^A)*(Z_B \ot \psi^B) = Z_A*
Z_B \ot \psi^A \psi^B$ is understood and the sum on the \rhs \
starts with $1 \ot 1$.
\subsection{The Hopf structure}
\label{THs}
We derive now the Hopf data for the new basis. 
A standard property of $\bfC$ is
\ble{copC}
( \Delta \ot \id) \bfC = \bfC_{13} \bfC_{23}
\, ,
\qquad 
\qquad
( \id \ot \Delta ) \bfC = \bfC_{12} \bfC_{13}
\, ,
\ee
where, \eg, $\bfC_{13} \equiv e^{\psi^A \ot 1 \ot Z_A}$ --- this
is just the product-coproduct duality in~(\ref{pdc}). The
second of~(\ref{copC}) permits the calculation of the coproduct
of the $\psi$'s by applying the Baker-Cambell-Hausdorff (BCH)
formula to the product on its \rhs, $\Delta(\psi^A)$ is the
coefficient of $Z_A$ in the resulting single exponential
\bae
\label{psicop}
\exp\bigl( Z_A \ot \Delta(\psi^A) \bigr) 
\fe
\exp \bigl( Z_A \ot \psi^A \ot 1 \bigr)
\exp \bigl( Z_B \ot 1 \ot \psi^B \bigr)
\ff
 \fe
\exp \Bigl( Z_A \ot \psi^A \ot 1 + Z_B \ot 1 \ot \psi^B
+ \frac{1}{2} [ Z_A, \, Z_B] \ot \psi^A \ot \psi^B + \dots
\Bigr)
\ff
\fe
\exp \Bigl\{ Z_A \ot \bigl( \psi^A \ot 1 + 1 \ot \psi^A + \frac{1}{2}
\f{B_1 B_2}{A} \psi^{B_1} \ot \psi^{B_2} + \dots 
\bigr)
\Bigr\}
\, , 
\eae
so that
\ble{coppsi}
\Delta(\psi^A) = 
\psi^A \ot 1 + 1 \ot \psi^A 
+ \frac{1}{2} \f{B_1 B_2}{A} \psi^{B_1} \ot \psi^{B_2} + \dots
\, .
\ee
Higher terms in the coproduct can be computed by using a
recursion relation for the BCH formula (see, \eg, Sec.{} 16
of~\cite{Bor.Rog.Sla:95}).
The counit of all $\psi^A$ vanishes.
Although $\Delta(\psi^A)$ can be complicated, $S(\psi^A)$ never
is. Using $\ip{S(g)}{\psi^A} = \ip{g}{S(\psi^A)}$ and the fact
that $S(g)=g^{-1}$, it is easily inferred that
\ble{Spsi}
S(\psi^A) = - \psi^A
\, ,
\ee
which extends as $S(p_r(\psi))= (-1)^r p_r(\psi)$ on homogeneous
polynomials of $p$-degree $r$. We see the first of the many
advantages of working in the $\psi$-basis: the antipode is diagonal. 
\begin{example}{Computation of $\psi^{(n)}$, $n \leq 4$}{CompAi}
A straightforward application of~(\ref{expexp}) gives
\bae
\label{phipsi}
\slte 
\Too 
\fe 
\slts 
\psi^{\Too} \ip{Z_{\Too}}{\slte \Too} 
\, = \, 
\psi^{\Too} 
\ff
\slte 
\Tto 
\fe
\slts \psi^{\Tto}  \ip{Z_{\Tto}}{\slte \Tto}
  + \frac{1}{2} \psi^{\Too} \psi^{\Too} 
    \ip{Z_{\Too} Z_{\Too}}{\slte \Tto}
\, = \, 
   \psi^{\Tto} + \frac{1}{2}  {\psi^{\Too}}^2
\ff
\slte \Ttho
\fe
\slts \psi^{\Ttho} + \psi^{\Too} \psi^{\Tto} + \frac{1}{6}
{\psi^{\Too}}^3
\ff
\slte \Ttht
\fe
\slts \psi^{\Ttht} + \psi^{\Too} \psi^{\Tto} + \frac{1}{3} 
{\psi^{\Too}}^3
\ff
\slte \Tfo
\fe
\slts \psi^{\Tfo} +  \psi^{\Too} \psi^{\Ttho} 
  + \frac{1}{2} {\psi^{\Tto}}^2 
  + \frac{1}{2} {\psi^{\Too}}^2 \psi^{\Tto}
  + \frac{1}{24} {\psi^{\Too}}^4
\ff
\slte \Tft
\fe
\slts 
\psi^{\Tft} + \psi^{\Too} \psi^{\Ttho} + \frac{1}{2} 
\psi^{\Too} \psi^{\Ttht} + \frac{2}{3} {\psi^{\Too}}^2
\psi^{\Tto} + \frac{1}{12} {\psi^{\Too}}^4
\ff
\slte \Tfth
\fe
\slts 
\psi^{\Tfth} + \frac{1}{2} \psi^{\Too} \psi^{\Ttho} 
+ \frac{1}{2} \psi^{\Too} \psi^{\Ttht} 
+ \frac{1}{2} {\psi^{\Tto}}^2 
+ \frac{5}{6} {\psi^{\Too}}^2 \psi^{\Tto} 
+ \frac{1}{8} {\psi^{\Too}}^4
\ff
\slte \Tff
\fe
\slts 
\psi^{\Tff}
+ \frac{3}{2} \psi^{\Too} \psi^{\Ttht}
+ {\psi^{\Too}}^2 \psi^{\Tto}
+ \frac{1}{4} {\psi^{\Too}}^4
\, .
\eae
Inverting the above expressions we find
\bae
\label{psiphi}
\slts 
\psi^{\Too}
\fe
\slte 
\Too
\ff
\slts
\psi^{\Tto}
\fe
\slte 
\Tto 
- \frac{1}{2} {\Too}^2
\ff
\slts 
\psi^{\Ttho}
\fe
\slte
\Ttho 
- \Too \Tto
+ \frac{1}{3} {\Too}^3
\ff
\slts
\psi^{\Ttht}
\fe
\slte
\Ttht 
- \Too \Tto
+ \frac{1}{6} {\Too}^3
\ff
\slts 
\psi^{\Tfo}
\fe
\slte
\Tfo 
- \Too \Ttho
-\frac{1}{2} {\Tto}^2 
+ {\Too}^2 \Tto
- \frac{1}{4} {\Too}^4
\ff
\slts
\psi^{\Tft}
\fe
\slte
\Tft 
- \Too \Ttho
+ \frac{5}{6} {\Too}^2 \Tto
-\frac{1}{2} \Too \Ttht
- \frac{1}{6} {\Too}^4
\ff
\slts
\psi^{\Tfth}
\fe
\slte
\Tfth
- \frac{1}{2} \Too \Ttho
- \frac{1}{2} \Too \Ttho
+ \frac{2}{3} {\Too}^2 \Tto
- \frac{1}{2} {\Tto}^2
- \frac{1}{12} {\Too}^4
\ff
\slts
\psi^{\Tff}
\fe
\Tff 
-\frac{3}{2} \Too \Ttht
+ \frac{1}{2} {\Too}^2 \Tto
\, .
\eae
\longpage
\longpage
Concerning the coproduct, Eq.~(\ref{coppsi}) shows that all ladder 
$\psi$'s are primitive. For
the rest of the $\psi$'s, we get (omitting the primitive part)
\bae
\label{copex}
\slts
\tilde{\Delta} \Bigl( \psi^{\Ttht} \Bigr) 
\fe
\slts
\psi^{\Too} \ot \psi^{\Tto} - \psi^{\Tto} \ot \psi^{\Too}
\ff
\slts
\tilde{\Delta} \Bigl( \psi^{\Tft} \Bigr) 
\fe
\slts
\psi^{\Too} \ot \psi^{\Ttho} - \psi^{\Ttho} \ot \psi^{\Too}
+\frac{1}{2} \psi^{\Ttht} \ot \psi^{\Too} 
-\frac{1}{2} \psi^{\Too} \ot \psi^{\Ttht}
\ff
 & & 
\slts
{} + \frac{1}{6} \psi^{\Too} \psi^{\Tto} \ot \psi^{\Too}
+ \frac{1}{6} \psi^{\Too} \ot \psi^{\Too} \psi^{\Tto}
- \frac{1}{6} {\psi^{\Too}}^2 \ot \psi^{\Tto}
- \frac{1}{6} \psi^{\Tto} \ot {\psi^{\Too}}^2
\ff
\slts
\tilde{\Delta} \Bigl( \psi^{\Tfth} \Bigr)
\fe
\slts
\frac{1}{2} \psi^{\Too} \ot \psi^{\Ttho}
-\frac{1}{2} \psi^{\Ttho} \ot \psi^{\Too}
+ \frac{1}{2} \psi^{\Too} \ot \psi^{\Ttht}
- \frac{1}{2} \psi^{\Ttht} \ot \psi^{\Too}
\ff
 & & 
\slts
{} - \frac{1}{6} \psi^{\Too} \ot \psi^{\Too} \psi^{\Tto} 
- \frac{1}{6} \psi^{\Too} \psi^{\Tto} \ot \psi^{\Too} 
+ \frac{1}{6} {\psi^{\Too}}^2 \ot \psi^{\Tto}
+ \frac{1}{6} \psi^{\Tto} \ot {\psi^{\Too}}^2 
\ff
\slts
\tilde{\Delta} \Bigl( \psi^{\Tff} \Bigr)
\fe
\slts
\frac{3}{2} \psi^{\Too} \ot \psi^{\Ttht}
- \frac{3}{2} \psi^{\Ttht} \ot \psi^{\Too}
- \frac{1}{2} \psi^{\Too} \ot \psi^{\Too} \psi^{\Tto}
- \frac{1}{2} \psi^{\Too} \psi^{\Tto} \ot \psi^{\Too}
\ff
 & &
\slts
{} + \frac{1}{2} {\psi^{\Too}}^2 \ot \psi^{\Tto}
+ \frac{1}{2} \psi^{\Tto} \ot {\psi^{\Too}}^2
\, .
\eae
One can easily verify that $S(\psi^A) = -\psi^A$.
\end{example}
\section{Primitive Elements}
\label{PE}
We turn now to the study of the primitive elements of $\calA$.
These are of fundamental importance in any Hopf algebra, but
acquire even more privileged status in our case, given their
r\^ole in renormalization. 
Apart from this, they are also of interest in representation
theory: given a primitive element $a \in \calA$, 
$\Delta(a)=a \otimes \uA +
\uA \otimes a$, one obtains a one-dimensional representation
$\rho_a$ of $\calU$ via 
\ble{odr}
\rho_a(x) \equiv \ip{x}{e^a}
\, .
\ee
Indeed, $e^a$ is group-like, $\Delta(e^a) =e^a \otimes e^a$, so
that 
\ble{multipl}
\rho_a(xy) \equiv \ip{xy}{e^a} = \ip{x \otimes y}{e^a \otimes e^a} 
  = \rho_a(x) \rho_a(y)
\, .
\ee
Conversely, every
one-dimensional representation of $\calU$ is associated to some
primitive element in $\calA$. 

Primitive elements are typically rare,
but the algebra of rooted trees is quite exceptional in this
respect:
there is an infinite number of them in $\calA$, with a non-trivial 
index set. 
We start our discussion with the easiest case, that of the ladder
generators, for which our Theorem~\ref{the:ladder} below supplies
a complete answer. We then turn to the considerably more 
complicated general
case which Theorem~\ref{the:gen} reduces to the problem of
finding all closed LI 1-forms on $G$.
\subsection{Ladder generators}
\label{Lg}
We consider the subalgebra $\calT$ of $\calH_R$ generated by
the ladder generators $T_n$, 
where $n$ counts the number of vertices. Their coproduct is
\ble{Tncop}
\Delta(T_n) = \sum_{k=0}^n  T_k \ot T_{n-k}
\, ,
\ee
making $\calT$ a sub-Hopf algebra of $\calH_R$ (notice though that for
$\phi$ not in $\calT$, $\Delta(\phi)$ may involve terms in $\calT
\ot \calT$). 
Experimenting a little we find that, for the first few $n$'s, 
each $T_n$
gives rise to a primitive $P^{(n)}$. The general case is
handled by the following
\begin{theorem}
\label{the:ladder}
To each ladder generator $T_{n}$, $n=1,2,\ldots$, corresponds a 
primitive element 
$P^{(n)}$, with $T_{n}$ as its linear part, given by
\fle{PnTn}{%
P^{(n)} = \frac{1}{n!} \frac{\partial^n}{\partial x^n} \log
\left( \sum_{m=0}^{\infty} T_m x^m \right) \bigg|_{x=0}
\, .%
}%
\end{theorem}
\begin{proof}
Consider the algebra $\calF$ of formal power series
$f(x)=\sum_{n=0}^{\infty} c_n x^n$, $c_0=1$, with the usual
product. Define a basis $\{\xi_n, \, n=0,1,2,\ldots \}$ of $\calF^*$, 
the dual of $\calF$, via
\ble{Tndef}
\ip{\xi_n}{f(x)} = c_n
\, ,
\ee
\ie, $\xi_n$ reads off the coefficient of $x^n$ in $f$ and
$\xi_0=1$. For $f''(x)
= f'(x) f(x)$ we have%
\footnote{Notice that primes only distinguish functions here, they 
do not denote differentiation.%
} 
\ble{fppffp}
f''(x) = \sum_{n=0}^{\infty} c_n'' x^n
\, , \qquad c_n'' = \sum_{k=0}^{n} c_k' c_{n-k}
\, ,
\ee
which implies the coproduct
\ble{coptn}
\Delta(\xi_{n}) = \sum_{k=0}^{n} \xi_k \ot \xi_{n-k}
\ee
in $\calF^*$. Endowing $\calF^*$ with a commutative product, we arrive
at the isomorphism $\calF^* \cong \calT$, as Hopf algebras, with $\xi_n
\leftrightarrow T_n$. Define a
new basis $\{ \sigma_n, \, n=0,1,2,\dots \}$ in $\calF^*$ by 
\ble{sin}
\ip{\sigma_r}{f(x)} = \tilde{c}_r
\, ,
\qquad \mbox{with} \quad 
f(x) = e^{\sum_{r=1}^\infty \tilde{c}_r x^r}
\ee
and $\sigma_0=1$. 
Then 
\ble{fppt}
f''(x) = e^{\sum_{r=1}^\infty \tilde{c}''_r x^r} 
\, , 
\qquad
\mbox{with}
\quad
\tilde{c}''_r = \tilde{c}'_r + \tilde{c}_r
\, ,
\ee
implying the coproduct $\Delta(\sigma_r) = \sigma_r \ot 1 + 1 \ot
\sigma_r$. The $\sigma$'s, under the above isomorphism, correspond
to the $P^{(n)}$ in $\calT$. Solving the equation
\ble{PnTneq}
e^{\sum_{r=1}^\infty P^{(r)} x^r} = \sum_{n=0}^\infty T_n x^n
\ee
for $P^{(r)}$, one arrives at~(\ref{PnTn}). 
\end{proof}
\noindent We read off $P^{(n)}$, for the first few values of
$n$,  as the coefficient of $x^n$ in the Taylor series expansion
\bae
\label{TTnPn}
\log \left( \sum_{n=0}^\infty T_n x^n \right) 
\fe
T_1 \, x
+ \big(T_2-\frac{1}{2} T_1^2\big) \, x^2
+ \big(T_3-T_1T_2+\frac{1}{3}T_1^3\big) \, x^3
\ff
\aqqa 
{}+\big(T_4-T_1 T_3 -\frac{1}{2} T_2^2
+ T_1^2 T_2 -\frac{1}{4} T_1^4\big) \, x^4
\ff
\aqqa 
{}+\big(T_5 - T_1 T_4 -T_2 T_3 + T_1^2 T_3 + T_1 T_2^2
-T_1^3 T_2 + \frac{1}{5} T_1^5\big) \, x^5
\ff
\aqqa 
{}+\big(T_6 -T_1 T_5 -T_2 T_4 -\frac{1}{2} T_3^2 
+ T_1^2 T_4 +2 T_1 T_2 T_3
-T_1^3 T_3 
\ff
\aqqa 
{} \phantom{+} + \frac{1}{3} T_2^3 - \frac{3}{2} T_1^2 T_2^2
+ T_1^4 T_2 - \frac{1}{6} T_1^6\big) \, x^6
+ \ldots 
\eae
The polynomials $P^{(n)}(T_i)$ are known as {\em Schur
polynomials}. 
\subsection{The general case}
\label{tgc}
Given a closed LI 1-form $\alpha$ on $G$, there exists 
a linear combination $\phi^{i'}$ of the generators $\phi^A$
such that $\alpha = \Pi_{\phi^{i'}}$. Applying the inverse
Poincar\'e lemma, we may write (locally) 
\ble{phiip}
\Pi_{\phi^{i'}}= d \psi^{i'}
\, ,
\ee
for some function $\psi^{i'}$ in $\calA$. Requiring additionally
that $\psi^{i'}$ vanish at the origin, $\epsilon(\psi^{i'}) =0$,
fixes the constant left arbitrary by~(\ref{phiip}) to zero.
$\psi^{i'}$ can be expressed in terms of the $\phi$'s. Since
$\Pi_{\phi^{i'}}$ reduces to $d \phi^{i'}$ at the origin, the
linear part $\psilin^{i'}$ of $\psi^{i'}(\phi)$ is $\phi^{i'}$. But then
$\Pi_{\phi^{i'}}=\Pi_{\psi^{i'}}$, since $\Pi$ projects to the
linear part. Comparing the \rhs \ of~(\ref{phiip}) with the general
expression for a LI 1-form, Eq.~(\ref{Pia}), we conclude that
$\psi^{i'}$ is primitive. Conversely, every primitive function
$\psi^{i'}$ gives rise to a closed LI 1-form, 
$d\Pi_{\psi^{i'}}=dd\psi^{i'}=0=d\Pi_{\psilin^{i'}}$. 

\noindent Eq.~(\ref{MCeq}), and the comment that follows it, show
that $\Dlin(\phi^{i'})$ is symmetric under the interchange of its
two tensor factors. This observation leads to a particularly
simple way to identify primitive elements. One first looks for
linear combinations $\phi^{i'}$ of the $\phi^A$ with symmetric
$\Dlin(\phi^{i'})$ (notice that $\Dlin$ is given by simple cuts).
The explicit expression for the corresponding primitive
$\psi^{i'}$ then is given by the standard formula for the (local)
potential of  
a closed form. We find that the result is simplified considerably
due to the particular form of the coproduct of the $\phi^A$,
namely the linearity of $\Delta(\phi^A)$ in its second tensor
factor. 
\begin{theorem}
\label{the:gen}
Given $\phi^{i'} \in \calA_1$, such that 
$d \Pi_{\phi^{i'}}=0$. Then the element $\psi^{i'}$ of  $\calA$,
given by
\fle{psiip}{%
\psi^{i'} = -\Phi^{-1} \circ S(\phi^{i'})
\, ,%
}
is primitive and has $\phi^{i'}$ as its linear part 
($\Phi$ above is the $p$-degree operator for the $\phi$'s,
$\Phi(\phi^{A_1} \dots \phi^{A_r}) = r \phi^{A_1} \dots
\phi^{A_r}$). 
\end{theorem}
\begin{proof}
We apply the inverse Poincar\'e lemma to
$\Pi_{\phi^{i'}}$. For a given $v$-degree $n$, only $\phi^A$ of
$v$-degree up to $n$ enter in the formulas --- we denote them
collectively by $\vec{x}$ (\eg, $S(\phi)(\vec{x})$ denotes the
standard expression of $S(\phi)$ in terms of the $\phi^A$ while
$S(\phi)(z\vec{x})$ denotes the same expression with every
$\phi^A$ multiplied by $z$). Consider the family of diffeomorphisms
$\varphi_t: \vec{x} \mapsto (1-t) \vec{x}$, $0 \leq t \leq 1$.
Then $\varphi^*_0$ is the identity map while $\varphi^*_1$ is the
zero map. The corresponding velocity field is
\ble{vecv}
\vec{v} 
\, = \, 
\frac{d}{dt} \varphi_t (\vec{x}) 
\, = \, 
- \vec{x} 
\, \Rightarrow \,
\vec{v}(\vec{y}, t) = -\frac{1}{1-t} \vec{y}
\, ,
\ee
where $\vec{y} = \varphi_t(\vec{x})$. We have%
\footnote{%
We ignore in the sequel the singularity of $\vec{v}$ at $t=1$ ---
it is easily  shown to be harmless.%
} 
\ble{Piphit}
\Pi_{\phi^{i'}}(\vec{x}) \, = \, 
\varphi^*_0 \bigl( \Pi_{\phi^{i'}} (\varphi_0(\vec{x})) \bigr)
 - \varphi^*_1 \bigl( \Pi_{\phi^{i'}} (\varphi_1(\vec{x})) \bigr) 
\, = \, 
\int_1^0 dt \,  \frac{d}{dt} \varphi^*_t \bigl( \Pi_{\phi^{i'}}
(\vec{y}) \bigr)
\, .
\ee
However, $\frac{d}{dt} \varphi^*_t = \varphi^*_t L_{\vec{v}} =
\varphi^*_t (d \, \ibf_{\vec{v}} + \ibf_{\vec{v}} \, d)$ 
and, taking into account the closure of $\Pi_{\phi^{i'}}$, we
find
\ble{Piphit2}
\Pi_{\phi^{i'}}(\vec{x}) 
\, = \, 
 d \, \int_1^0 dt \, \varphi^*_t \bigl( \ibf_{\vec{v}} \, 
\Pi_{\phi^{i'}} (\vec{y}) \bigr)
\, .
\ee
This is the inverse Poincar\'e lemma. 
We concentrate now on the action of $\ibf_{\vec{v}}$ on
$\Pi_{\phi^{i'}} (\vec{y})$. We have 
\ble{acti}
\ibf_{\vec{v}} =
-\frac{1}{1-t} \, y^j \, \ibf_{\partial_{ y^j}}
\, , 
\qquad \qquad
\Pi_{\phi^{i'}} (\vec{y}) = S(\phi^{i'}_{(1)}) \, d
\phi^{i'}_{(2)}(\vec{y})
\, .
\ee
In this latter (implied) sum, all terms in the
coproduct of $\phi^{i'}$ appear except the first one, $\phi^{i'}
\ot 1$, which is annihilated by $d$. Notice now that $y^j \, 
\ibf_{\partial_{y^j}} dy^i = y^i$. Since
$\Delta(\phi^{i'})$ is linear in its second factor we conclude
that 
\ble{ivPi}
y^j \ibf_{\partial_{y^j}}
S(\phi^{i'}_{(1)}) \, d \phi^{i'}_{(2)}(\vec{y}) 
\, = \,
S(\phi^{i'}_{(1)}) \, \phi^{i'}_{(2)}(\vec{y}) -
S(\phi^{i'})(\vec{y}) 
\, = \,
- S(\phi^{i'})(\vec{y})
\, .
\ee
Substituting back into~(\ref{Piphit2}) and putting $1-t \equiv z$
we find
\ble{Piphit3}
\Pi_{\phi^{i'}}(\vec{x})
\, = \,
- d \, \int_0^1 \frac{dz}{z} \, S(\phi^{i'})(z \vec{x})
\, ,
\ee
which, upon performing the integration over $z$,
gives $\Pi_{\phi^{i'}} = - d \Phi^{-1} \circ S(\phi^{i'})$. The
remarks preceding the theorem complete the proof. 
\end{proof}
\subsection{The lower central series and $k$-primitiveness}
\label{lcskp}
We extend here the notion of primitiveness to that of
$k$-primitiveness. Our starting point is our BCH-based
prescription for
calculating the coproduct of the $\psi$'s, Eq.~(\ref{coppsi}). 
Suppose we identify all generators $Z^{[1]}_{i}$ of $\calG$
that cannot be written as commutators
(the $Z^{[1]}_{i}$ are, in general, linear combinations 
of the $Z_{A}$). 
Then we may
perform a linear change of basis in $\calG$ and split the
generators into two classes, one made up of the above
$Z^{[1]}_{i}$
and the other spanning the 
complement --- we denote the latter by $\{ Z_{i'} \} $. 
Writing the canonical element  in the new basis,
\ble{Cpsip}
\bfC = 
 e^{ \psi^i_{[1]} \ot Z_i^{[1]}
  + \psi^{i'} \ot Z_{i'} }
\, ,
\ee
we are led to the identification of the $\psi^{i}_{[1]}$ with the 
primitive
$\psi$'s. This is so since, in the BCH formula, the $Z_{i}^{[1]}$ are
never produced by the commutators, so that the only contribution
to $\Delta(\psi^{i}_{[1]})$ is the primitive part. 
Consider now the lower central series of $\calG$,
consisting of the series of subspaces $\calG^{[1]}$,
$\calG^{[2]}, \dots$.
A particular $Z$
in $\calG$ belongs to $\calG^{[k]}$ if it can be written as a
$(k-1)$-nested commutator. This implies that if $Z$ belongs to
$\calG^{[k]}$, it also belongs to all $\calG^{[r]}$, with $r <
k$. This is the standard definition of $\calG^{[k]}$ --- we 
actually need a slightly modified
one, according to which {\em $Z$ belongs only to the $\calG^{[k]}$
with the maximum $k$}. With this definition, $\calG^{[k]} \cap
\calG^{[r]} = \emptyset$ whenever $k \neq r$. We may now 
perform a linear
change of basis in $\calG$ such that each generator $Z^{[k]}_i$
in the new basis belongs to $\calG^{[k]}$. Writing the canonical
element in the form 
\ble{CGk}
\bfC = e^{ \psi^i_{[k]} \ot Z^{[k]}_i}
\, ,
\ee
defines the {\em $k$-primitiveness} for the $\psi^i_{[k]}$ dual to the
above $Z^{[k]}_i$. Since the $Z^{[k]}_i$ are linear combinations
of the $Z_A$, the $\psi^i_{[k]}$ will be linear combinations of
the $\psi^A$. $\calA$ splits accordingly to a direct sum, $\calA
=\bigoplus_{k=1}^{\infty} \calA^{[k]}$ --- the primitive
$\psi$'s, in particular, span $\calA^{[1]}$. Notice that $\psi$'s
with $n$ vertices may belong to $\calG^{[k]}$ with $k \leq n-1$.
This is so because the ``longest'' nested commutator with $n$
vertices is $\slts [Z_{\Too}, \, [ Z_{\Too}, \dots \, 
[ Z_{\Too}, \, Z_{\Tto}
]] \dots ]$, with $n-2$ entries of $\slts Z_{\Too}$. 

The above concept of $k$-primitiveness arose naturally in our
study of the primitive $\psi^i_{[1]}$. Some time afterwards,
we became aware of Ref.~\cite{Bro.Kre:00}, where a
concept of $k$-primitiveness is also defined, as follows:
given an element $\chi$ of $\calA$, one computes successive
powers of the coproduct, $\Delta^k(\chi)$. There is a minimum $k$
for which all terms in $\Delta^k(\chi)$ contain a unity in at
least one of the tensor factors --- this defines the
$k$-primitiveness of $\chi$. Our
definition is intrinsically defined only on the generators
$\psi^i_{[k]}$, while the above makes sense in all of $\calA$. We
now show that, for $\psi^i_{[k]}$, the two definitions coincide. 
\begin{lemma}
\label{Dkmin}
The minimum value of $r$ for which $\Delta^r(\psi^i_{[k]})$ 
contains at least one unit tensor factor in
each of its terms, is $r=k$.
\end{lemma}
\begin{proof}
The various powers of the coproduct of $\psi^i_{[k]}$ can be
computed by iteration of the second of~(\ref{copC}),
\ble{Dkpsi}
\Delta^{r-1}(\psi^i_{[k]}) = \text{coeff. of \ } Z^{[k]}_i 
\text{ \ in \ } 
\log \Bigl ( \bfC_{01} \bfC_{02} \dots \bfC_{0r} \Bigr )
\, .
\ee
This shows that in $\Delta^k(\psi^i_{[k]})$, the $(k+1)$-linear
term can only be produced by the $k$-nested commutator 
\[ [ Z_{i_1}, \, [ Z_{i_2}, \, \dots [ Z_{i_k}, \, Z_{i_{k+1}} ]]
\dots ] \ot \psi^{i_1} \ot \dots \ot \psi^{i_{k+1}} 
\, .
\]
The latter, however, has no $Z^{[k]}_i$ component, since
$Z^{[k]}_i$ can
be written as a $(k-1)$-nested commutator at most. It is also
clear, for the same reason, that there are no terms of higher
$p$-degree in the $\psi$'s, as those would correspond to even
longer nested commutators. $\Delta^k(\psi^i_{[k]})$ then must
have at least one unit tensor factor in each of its terms. 
On the other hand, 
the $k$-linear term
in $\Delta^{k-1}(\psi^i_{[k]})$ is not zero, because, by
definition, the corresponding $(k-1)$-nested commutator has a
$Z^{[k]}_i$ component.
\end{proof}
As shown in~\cite{Bro.Kre:00}, the $k$-degree satisfies
\ble{kdegprop}
\text{deg}_k(\psi^i_{[k_1]} \psi^j_{[k_2]}) = k_1 + k_2
\, .
\ee
We use the two definitions of the $k$-degree interchangeably in
what follows. 
We may now clarify the relation between the primitive elements given
by the inverse Poincar\'e formula, Eq.~(\ref{psiip}), and the ones 
introduced above via the lower central series of $\calG$. 
\begin{lemma}
\label{phicphi}
Given $\phi^{i'} = {c^{i'}}_{ \! \! A} \phi^A$, with
${c^{i'}}_{\!\!  A}$
constants, such that $d \Pi_{\phi^{i'}}=0$. Then the primitive
element $\psi^{i'}$ of~(\ref{psiip}) is equal to ${c^{i'}}_{\! \! A}
\psi^A$, \ie,
\ble{psitw}
\psi^{i'} = -\Phi^{-1} \circ S (\phi^{i'}) = {c^{i'}}_{\! \! A} \psi^A
\, .
\ee
All primitive elements of $\calA$ can be obtained in this form.
\end{lemma}
\begin{proof}
Any linear combination of the $\psi^{i}_{[1]}$ is primitive,
while (sums of) products of them are not, due to~(\ref{kdegprop}). 
Therefore,
the $\psi^{i}_{[1]}$ constitute a linear basis in the vector space of
primitive elements of $\calA$. To the given $\phi^{i'}$,
Eq.~(\ref{psiip}) associates a primitive element $\psi^{i'}$, 
with $\phi^{i'}$ as its linear part.
The unique linear combination of the $\psi^A$ (and,
hence, of the $\psi^{i}_{[1]}$) with this linear part is
$\psi^{i'} = {c^{i'}}_{\! \! A} \psi^A$.
\end{proof}
We give an example illustrating the above.
\begin{example}{Construction of $\calG^{(n)[k]}$,
$\calA^{(n)[k]}$, for $n \leq 4$}{compPn}
To identify the generators of $\calG^{(n)[k]}$, we construct all
$(k-1)$-nested commutators with $n$ vertices --- $\calG^{(n)[1]}$
is determined as the complement of $\bigoplus_{k=2}^{n-1}
\calG^{(n)[k]}$ in $\calG^{(n)}$ (below we use the orthogonal
complement but this is not essential, one simply has to complete
the basis of the $Z$'s). This gives a
matrix that effects the transition from the basis $\{ Z_A \}$,
indexed by rooted trees, to the basis $\{ Z^{[k]}_i \}$, of
definite $k$-primitiveness.  The inverse matrix then gives the
$\psi^i_{[k]}$ in terms of the $\psi^A$. 

\noindent $\calG^{(1)[1]} = \calG^{(1)}$ is generated by 
$\slts Z_{\Too}$. $\calG^{(2)[1]} = \calG^{(2)}$ is generated by
$\slts Z_{\Tto}$, since the only commutator with two vertices,
$\slts [Z_{\Too}, \, Z_{\Too} ]$ is zero. For $n=3$, we have the
only non-zero commutator%
\footnote{We remind the reader our notation: $Z_i^{(n)[k]}$ is
the $i$-th element in the subspace of $k$-primitive, $n$-vertex
$Z$'s. The same notation is used for the $\psi$'s, with the
position of the indices (upper--lower) interchanged.%
}
$\slts Z^{(3)[2]}_1 \equiv [Z_{\Too}, \,
Z_{\Tto} ] = 2 Z_{\Ttht}$. The complement in $\calG^{(3)}$ is
spanned by $\slts Z^{(3)[1]}_1 = Z_{\Ttho}$. 
Next we look at the case $n=4$.
We find the only non-zero commutators
\ble{neq4}
\slts
[Z_{\Too}, \, Z_{\Ttho}] = (0,2,1,0) \equiv Z^{(4)[2]}_1
\, ,
\qquad \qquad
[Z_{\Too}, \, Z_{\Ttht}] = (0,-1,1,3) \equiv Z^{(4)[3]}_1
\, ,
\ee
in the basis 
$\sltss \bigl \{ Z_{\Tfo}, \, Z_{\Tft}, \, Z_{\Tfth}, \, Z_{\Tff}
\bigr \}$.
The orthogonal complement in $\calG^{(4)}$ is
spanned by 
\ble{Gfo}
Z^{(4)[1]}_1 = (1,0,0,0)
\, , 
\qquad \qquad
Z^{(4)[1]}_2 = (0,1,-2,1)
\, .
\ee 
Writing the above change of basis symbolically as $Z^{[k]}_i = M
Z_{A}$, with $M$ a matrix of numerical coefficients, the dual
change of basis for the $\psi$'s is given by $\psi_{[k]}^i =
\psi^A M^{-1}$. We find
\ble{psinew}
\slts 
\psi_{(1)[1]}^1 = \psi^{\Too}
\, ,
\qquad
\quad
\psi_{(2)[1]}^1 = \psi^{\Tto}
\, ,
\qquad
\quad
\psi_{(3)[1]}^1 = \psi^{\Ttho}
\, ,
\qquad
\quad
\psi_{(3)[2]}^1 = \frac{1}{2} \psi^{\Ttht}
\, ,
\ee
while, for $n=4$,
\bae
\slts
\psi_{(4)[1]}^1 
\fe
\slts
\psi^{ \Tfo}
\, ,
\ff
\psi_{(4)[1]}^2
\fe
\slts
\frac{1}{6} \psi^{\Tft} - \frac{1}{3} \psi^{\Tfth} 
+ \frac{1}{6} \psi^{\Tff}
\, ,
\ff
\psi_{(4)[2]}^1
\fe
\slts
\frac{7}{18} \psi^{\Tft} + \frac{2}{9} \psi^{\Tfth} + \frac{1}{18}
\psi^{\Tff}
\, ,
\ff
\psi_{(4)[3]}^1
\fe
\slts
- \frac{1}{18} \psi^{\Tft} + \frac{1}{9} \psi^{\Tfth} +
\frac{5}{18} \psi^{\Tff}
\, .
\eae
Referring to, \eg, $\psi^{(4)[1]}_2$, one easily verifies that
\ble{phiipex}
\slts
\phi^{i'} = \frac{1}{6} \phi^{\Tft} - \frac{1}{3} \phi^{\Tfth} +
\frac{1}{6} \phi^{\Tff}
\ee
has symmetric $\Dlin$ and, when inserted in~(\ref{psiip}),
delivers $\psi^{(4)[1]}_2$.
\end{example}
To continue the above construction to the cases $n=5,6$, we  
developed a REDUCE program, incorporating some of the
procedures of~\cite{Bro.Kre:99}.
The numbers $P_{n,k}$ of $k$-primitive $\psi$'s with $n \leq 6$ 
vertices
that we find coincide with the ones in Table 4
of~\cite{Bro.Kre:00}, as expected. In what refers to the
primitive $\psi$'s, the procedure presented above,
starting with $\phi$'s with symmetric $\Dlin$ and then
using~(\ref{psiip}), 
should be considerably more efficient than the one used
in~\cite{Bro.Kre:00} --- it would be interesting to quantify this 
statement. Notice that an equivalent procedure involves expanding
the primitive $\psi$'s as $\psi_{[1]}= c_A\psi^A$ and then
determining the constants $c_A$ from the set of equations
$\f{RS}{T} \ip{Z_T}{\psi_{[1]}}=0$ (the latter is the statement
that $\psi_{[1]}$ is invariant under the coadjoint coaction). 
\section{Normal Coordinates and Toy Model Renormalization}
\label{NCR}
We turn now to what, in some sense, is our main objective,
namely, the application of the formalism presented so far in the
problem of renormalization in perturbative quantum field theory.
The scope of our considerations in this section can only be
modest, since realistic quantum field theories involve rooted
trees with an infinite number of decorations. Nevertheless, a toy
model exists (see~\cite{Kre:98}) that realizes the $\phi^A$ as
nested divergent integrals, regulated by a parameter $\epsilon$.
We find this an extremely useful construct that captures many of
the most important features of realistic renormalization --- again,
we refer the reader to~\cite{Kre:98, Con.Kre:98} for a detailed
presentation. What we are interested in here, is the r\^ole of
the new coordinates $\psi$ in the renormalization of divergent
quantities. We start with a brief review of the basics. 
\subsection{The toy model}
\label{Tm}
The elementary divergence in the toy model we deal with is given
by the integral
\ble{elemdiv}
I_1(c; \, \epsilon) = \int_0^\infty dy \frac{y^{-\epsilon}}{y+c}
\, ,
\ee
which diverges as $\epsilon$ goes to zero. $c$ above will be
referred to as the {\em external parameter} of the integral. 
We associate the
function $\slts \phi^{\Too}$ with $I_1(c; \, \epsilon)$. 
To the function $\slts \phi^{\Tto}$ corresponds the nested integral
\ble{ttoint}
I_2(c; \, \epsilon) = 
\int_0^\infty dy_1 \frac{y_1^{-\epsilon}}{y_1+c} I_1(y_1; \, \epsilon)
=
\int_0^\infty dy_1 \frac{y_1^{-\epsilon}}{y_1+c}
\int_0^\infty dy_2 \frac{y_2^{-\epsilon}}{y_2+y_1}
\, .
\ee
Notice that the external parameter of the subdivergence $I_1$ is
$y_1$.
To $\slts \phi^{\Ttho}$, $\slts \phi^{\Ttht}$ correspond,
respectively, 
\ble{tthotint}
I_{3,1}(c; \, \epsilon) = 
\int_0^\infty dy_1 \, \frac{y_1^{-\epsilon}}{y_1+c} \, 
I_2(y_1; \, \epsilon)
\, ,
\qquad
\quad
I_{3,2}(c; \, \epsilon) =
\int_0^\infty dy_1 \,  \frac{y_1^{-\epsilon}}{y_1+c}  \, 
\bigl ( I_1(y_1; \, \epsilon) \bigr )^2
\, ,
\ee
it should be clear how this assignment extends to all $\phi^A$. 
In this way, each $\phi^A$ can be associated with the  Laurent
series in $\epsilon$ that corresponds to its associated integral, \eg
\ble{Lse}
\slts 
\phi^{\Too} 
= 
\int_0^\infty dy \frac{y^{-\epsilon}}{y+c} 
= 
\frac{\pi}{\sin(\pi \epsilon)} c^{-\epsilon} 
= 
\frac{1}{\epsilon} -a + \calO(\epsilon)
\, ,
\ee
where $a \equiv \log(c)$ and, similarly (using MAPLE),
\bae
\label{Lsexp}
\slts
\phi^{\Tto} 
\fe
\frac{1}{2\epsilon^2} - \frac{a}{\epsilon} + a^2 + \frac{5\pi^2}{12}
+ \calO(\epsilon)
\ff
\slts
\phi^{\Ttho}
\fe
\frac{1}{6 \epsilon^3} -\frac{a}{2 \epsilon^2} 
+ \Bigl( \frac{3a^2}{4} + \frac{7 \pi^2}{18}
\Bigr)\frac{1}{\epsilon}
- \frac{a}{12} \bigl( 9 a^2 + 14 \pi^2 \bigr)
+ \calO(\epsilon)
\ff
\slts
\phi^{\Ttht}
\fe
\frac{1}{3\epsilon^3} -\frac{a}{\epsilon^2}
+ \Bigl( 
\frac{3a^2}{2} + \frac{11\pi^2}{18} 
\Bigr) \frac{1}{\epsilon}
- \frac{a}{6}
\bigl(
9 a^2 + 11 \pi^2
\bigr)
+ \calO(\epsilon)
\ff
\slts
\phi^{\Tfo}
\fe
\frac{1}{24 \epsilon^4} 
- \frac{a}{6 \epsilon^3}
+ \Bigl( \frac{a^2}{3} + \frac{5 \pi^2}{24} \Bigr)
\frac{1}{\epsilon^2}
- \frac{a}{18} \Bigl( 8 a^2 + 15 \pi^2 \Bigr) \frac{1}{\epsilon}
+ \calO(\epsilon^0)
\\
\slts
\phi^{\Tft}
\fe
\frac{1}{12 \epsilon^4} 
- \frac{a}{3 \epsilon^3}
+ \Bigl( \frac{2 a^2}{3}  + \frac{3 \pi^2}{8} \Bigr)
\frac{1}{\epsilon^2}
- \frac{a}{18} \Bigl( 16 a^2 + 27 \pi^2 \Bigr) \frac{1}{\epsilon}
+ \calO(\epsilon^0)
\ff
\slts
\phi^{\Tfth}
\fe
\frac{1}{8 \epsilon^4} 
- \frac{a}{2 \epsilon^3}
+ \Bigl( a^2 + \frac{11 \pi^2}{24} \Bigr)
\frac{1}{\epsilon^2}
- \frac{a}{6} \Bigl( 8 a^2 + 11 \pi^2 \Bigr) \frac{1}{\epsilon}
+ \calO(\epsilon^0)
\ff
\slts
\phi^{\Tff}
\fe
\frac{1}{4 \epsilon^4} 
- \frac{a}{\epsilon^3}
+ \Bigl( 2 a^2 + \frac{19 \pi^2}{24} \Bigr)
\frac{1}{\epsilon^2}
- \frac{a}{6} \Bigl( 16 a^2 + 19 \pi^2 \Bigr) \frac{1}{\epsilon}
+ \calO(\epsilon^0)
\nonumber
\, ,
\eae
and so on. It is easily seen that $\phi$'s with $n$ vertices
give rise to Laurent series with leading pole of order $n$. 
The process of renormalization assigns to each $\phi^A$ a
finite ``renormalized'' value $\phi^A_\calR$ (see, \eg,~\cite{Col:84}).
In Hopf algebraic terms, the latter is given by~\cite{Bro.Kre:99}
\ble{phiR}
\phi^A_\calR = S_\calR \bigl( \phi^A_{(1)} \bigr) \phi^A_{(2)}
\, ,
\ee
where the {\em twisted antipode} $S_\calR$ is defined recursively
by
\ble{tSdef}
S_\calR \bigl( \phi^A \bigr) = - \calR \bigl( \phi^A \bigr) 
- \calR \Bigl( S_\calR \bigl( \phi^A_{(1')} \bigr) \phi^A_{(2')}
\Bigr)
\, .
\ee
$\calR$ above is a {\em renormalization map} that we choose here
to give the pole part of its argument, evaluated at the external
parameter equal to 1, \eg, $\slts \calR \bigl( \phi^{\Tto} \bigr) =
1/2\epsilon^2$ (compare with the first of~(\ref{Lsexp})). 
The primed sum in the second term
of~(\ref{tSdef}) excludes the primitive part of the coproduct. 
The magic of renormalization lies in the fact that, for any
$\phi^A$, the renormalized $\phi^A_\calR$ in~(\ref{phiR}) has no
poles in $\epsilon$ --- what makes this statement non-trivial is
that all terms subtracted iteratively from $\phi^A$, to give
$\phi^A_\calR$, are
independent of external parameters. We conclude our brief review 
with the following statement, proven in~\cite{Kre:99}: if $\calR$
satisfies the {\em multiplicative constraint}
\ble{multcon}
\calR (xy) - \calR \bigl( \calR(x)y \bigr) 
- \calR \bigl( x\calR(y) \bigr) + \calR(x) \calR(y) =0
\, ,
\ee
then $S_\calR$ is multiplicative, $S_\calR(xy)= S_\calR(x)
S_\calR(y)$ --- our choice of $\calR$ above does
satisfy~(\ref{multcon}). 
\subsection{Renormalization in the $\psi$-basis}
\label{Ritpb}
For a given number $n$ of vertices, the renormalization of every 
generator $\phi^A$ gives rise to $2^n$ counterterms, for a total
of $r_n 2^n$, where $r_n$ is the number of rooted trees with $n$
vertices. To renormalize the $\psi$'s, one can always express
them in terms of the $\phi$'s and then proceed as above. However,
for renormalization schemes $\calR$ that satisfy~(\ref{multcon}),
a much more efficient possibility arises. Eq.~(\ref{phiR}), in
this case, is valid for {\em any} function in $\calA$, and, in
particular, for the $\psi$'s. Notice that although the action of
the antipode
$S$ is trivial on the $\psi^A$, that of the twisted antipode
$S_\calR$ is not, in general. The advantage of working in the
basis $\{ \psi^i_{[k]} \}$ is that  the complexity of the 
renormalization of a generator $\psi^i_{(n)[k]}$ is governed by
$k$, not $n$, which entails, in general, significant savings. 
As an extreme example, a primitive $\psi$ with one hundred
vertices is renormalized by a simple subtraction --- this should
be compared with the $2^{100}$ counterterms necessary for the
renormalization of each of the $\phi_{(100)}$'s. 
How significant can 
the savings be in, \eg, CPU time,
depends on the distribution of the $\psi^i_{(n)}$ in the various
$k$-classes. As proved
in~\cite{Bro.Kre:00}, the numbers $P_{n,k}$ of $k$-primitive
$\psi$'s with $n$ vertices are generated by
\ble{Pnkgen}
P_k(x) \equiv \sum_{n=1}^\infty P_{n,k} x^n 
= \sum_{s|k} \frac{\mu(s)}{k} \Bigl( 
1-\prod_{n=1}^\infty \bigl( 1-x^{ns} \bigr)^{r_n}
\Bigr)^{k/s}
\, ,
\ee
a rather non-trivial result. The sum in the \rhs \ above extends
over all divisors $s$ of $k$, including $1$ and $k$. $\mu(s)$ is
the M\"obius function, equal to zero, if $s$ is divisible
by a square, and to $(-1)^p$, if $s$ is the product of $p$
distinct primes $(\mu(1) \equiv 1)$. 
Of particular interest to us is
the asymptotic behavior of $P_{n,k}$, for large values of
$n$~\cite{Bro.Kre:00}
\ble{asyPnk}
f_k \equiv \lim_{n \to \infty} \frac{P_{n,k}}{r_n} = 
\frac{1}{c} \, 
\Bigl( 
1 - \frac{1}{c} \Bigr)^{k-1}
\, ,
\ee
where $c=2.95\dots$ is the Otter constant. This is 
encouraging, as the population of the CPU-intensive high-$k$
$\psi$'s is seen to be exponentially suppressed. A
realistic estimate of the complexity of  renormalization in
the $\psi$-basis is outside the scope of this article, as it
would probably entail implementation-dependent parameters.
Nevertheless, we attempt a first-order estimation by assigning a 
computational cost of $2^k$ to a $k$-primitive $\psi$, while the
$\phi_{(n)}$ are assigned the cost $2^n$. The ratio of 
the total costs of renormalizing 
all generators with $n$ vertices in the two bases then is 
\ble{relco}
\rho_n = \frac{r_n 2^n}{\sum_{k=1}^{n-1} P_{n,k} 2^k} 
\approx (c-2) \Bigl( \frac{c}{c-1} \Bigr)^{n-1}
\, ,
\ee
with $\rho_{33} \approx 6 \times 10^5$ making
the difference between a week and a second.
We consider~(\ref{relco}) as
a loose upper bound on the potential savings. 

Another feature of the $\psi$'s that is worth pointing out is
their toy model pole structure. As mentioned above, each of the
$\phi^A_{(n)}$ corresponds to a Laurent series with maximal pole
order $n$. We find that the behavior of the $\psi^i_{(n)}$ is
much milder. We list the series expansion of the first few
$\psi^A$, which should be compared with the analogous expressions
for the $\phi^A$, Eq.~(\ref{Lsexp}),
\bae
\label{psiser}
\slts
\psi^{\Too} 
\fe
\frac{1}{\epsilon} -a + \calO(\epsilon)
\ff
\slts
\psi^{\Tto} 
\fe
\frac{\pi^2}{4}
+ \calO(\epsilon)
\ff
\slts
\psi^{\Ttho}
\fe
\frac{\pi^2}{18 \epsilon}
-\frac{\pi^2 a}{6}
+ \calO(\epsilon)
\ff
\slts
\psi^{\Ttht}
\fe
\frac{7 \pi^2}{36 \epsilon}
-\frac{7 \pi^2 a}{12}
+ \calO(\epsilon)
\ff
\slts
\psi^{\Tfo}
\fe
\frac{\pi^4}{8}
+ \calO(\epsilon)
\\
\slts
\psi^{\Tft}
\fe
\frac{19 \pi^4}{72}
+ \calO(\epsilon)
\ff
\slts
\psi^{\Tfth}
\fe
\frac{\pi^2}{24 \epsilon^2}
- \frac{\pi^2 a}{6\epsilon}
+ \calO(\epsilon^0)
\ff
\slts
\psi^{\Tff}
\fe
\frac{\pi^2}{12 \epsilon^2}
- \frac{\pi^2 a}{3 \epsilon}
+ \calO(\epsilon^0)
\nonumber
\, .
\eae
Notice that, \eg, the
primitive $\sltss \psi^{\Tfo}$ is actually finite, as is $\sltss
\psi^{\Tft}$ which is not primitive. We emphasize that $\sltss
\Bigl( \psi^{\Tft} \Bigr)_\calR$ is still given by~(\ref{phiR})
(with $\phi^a \to \sltss \psi^{\Tft}$) and does not 
coincide with
the finite $\sltss \psi^{\Tft}$ (see Ex.{}~\ref{Renorm} below). 
The other two $\psi_{(4)}^i$ are
of order $1/\epsilon^2$, even though they have $\calG^{[3]}$
components. These initial observations point to a general feature
of the $\psi$'s: the pole order 
does not specify the complexity of their renormalization, as is the
case with the $\phi$'s. The cancellations of the
higher-order poles observed point to rather non-trivial
underlying combinatorics that, we believe, deserve further
investigation.  

\noindent The series expansion of the $\psi^i_{(n)[k]}$ is
\bae
\label{serpsink}
\psi^2_{(4)[1]} 
\fe
\frac{\pi^4}{48}
+ \calO(\epsilon)
\ff
\psi^1_{(4)[2]} 
\fe
\frac{\pi^2}{72 \epsilon^2}
-\frac{\pi^2 a}{18 \epsilon}
+ \calO(\epsilon^0)
\\
\psi^1_{(4)[3]} 
\fe
\frac{\pi^2}{36 \epsilon^2}
-\frac{\pi^2 a}{9 \epsilon}
+ \calO(\epsilon^0)
\nonumber
\eae
(the rest are essentially identical to the $\psi^A$).
We also point out that some of the $n=6$
primitive $\psi$'s are of order $1/\epsilon^3$ --- nevertheless,
the coefficients of all poles are independent of $c$ and their
renormalization is accomplished by a simple subtraction, in
agreement with~(\ref{phiR}).
\begin{example}{Renormalization of $\psi^2_{(4)[1]}$, 
$\psi^1_{(4)[2]}$, $\slts \psi^{\Tft}$}{Renorm}
For the primitive $\psi^2_{(4)[1]}$,
Eqs.~(\ref{tSdef}),~(\ref{serpsink}) give 
\ble{Scompex}
S_\calR \bigl( \psi^2_{(4)[1]} \bigr) 
= - \calR \bigl( \psi^2_{(4)[1]} \bigr)
= 0
\, ,
\ee
so that the renormalized value $ \bigl( \psi^2_{(4)[1]}
\bigr)_\calR = \psi^2_{(4)[1]} + S_\calR \bigl( \psi^2_{(4)[1]} \bigr)$ 
coincides with $\psi^2_{(4)[1]}$.
For the 2-primitive $\psi^1_{(4)[2]}$, the first of~(\ref{neq4}) 
and~(\ref{coppsi}) give
\ble{coppsi421}
\Delta \bigl( \psi^1_{(4)[2]} \bigr)
=
\psi^1_{(4)[2]} \ot 1 + 1 \ot \psi^1_{(4)[2]} 
+ \frac{1}{2} \psi^1_{(1)[1]} \ot \psi^1_{(3)[1]} 
- \frac{1}{2} \psi^1_{(3)[1]} \ot \psi^1_{(1)[1]} 
\, ,
\ee
so that 
\ble{psirenex2}
\bigl( \psi^1_{(4)[2]} \bigr)_\calR
=
\psi^1_{(4)[2]} + S_\calR \bigl( \psi^1_{(4)[2]} \bigr)
+ \frac{1}{2} S_\calR \bigl( \psi^1_{(1)[1]} \bigr)
\psi^1_{(3)[1]}
- \frac{1}{2} S_\calR \bigl( \psi^1_{(3)[1]}  \bigr)
\psi^1_{(1)[1]}
\, .
\ee
For the (non-trivial) twisted antipode we find
\ble{tSpsiex}
S_\calR \bigl(  \psi^1_{(4)[2]} \bigr)
= - \calR \bigl( \psi^1_{(4)[2]} \bigr)
+ \frac{1}{2} \calR \Bigl( \calR \bigl( \psi^1_{(1)[1]} \bigr)
\psi^1_{(3)[1]} \Bigr)
- \frac{1}{2} \calR \Bigl( \calR \bigl( \psi^1_{(3)[1]} \bigr)
\psi^1_{(1)[1]} \Bigr)
\, .
\ee
Substituting above we get
\ble{psi421ren}
\bigl( \psi^1_{(4)[2]} \bigr)_\calR
= \frac{7}{96} \pi^4 + \calO(\epsilon)
\, .
\ee
Finally, for $\slts \psi^{\Tft}$, we use the coproduct given
in~(\ref{copex}) and, proceeding along the same lines, we find
\ble{pftr}
\bigl( \psi^{\slts \Tft} \bigr)_\calR
=
\frac{13}{96} \pi^4 - \frac{1}{24} \pi^2 a^2 + \calO(\epsilon)
\, ,
\ee
which is different, as mentioned above, from the finite
$\psi^{\slts \Tft}$.
\end{example}
The remarkable pole structure of the $\psi$'s observed above,
persists for other, more realistic models as well. For example, 
we have repeated the above analysis for the heavy-quark model
of~\cite{Bro.Kre:99}. We find that, for $n \leq 4$, the maximal
pole order appearing is only $1/\epsilon$, with all ladder
$\psi$'s, except the first one, finite.
\section*{Acknowledgements}
C.{} C.{} would like to thank Denjoe O'Connor for discussions and
for pointing out
Ref.~\cite{Mil:84}.
The authors acknowledge partial support from CONACyT grant
32307-E and DGAPA-UNAM grant IN119792 (C.{} C.), DGAPA-UNAM grant
981212 (H.{} Q.) and CONACyT project G245427-E (M.{} R.).

\end{document}